# Student Understanding of the Bloch Sphere

Peter Hu*, Yangqiuting Li, Roger S. K. Mong, and Chandralekha Singh

*Department of Physics and Astronomy, University of Pittsburgh, Pittsburgh, PA 15260*
*Corresponding Author, email: pth9@pitt.edu

**Abstract**

Quantum information science is a rapidly growing interdisciplinary field that is attracting the attention of academics and industry experts alike. It requires talent from a wide variety of traditional fields, including physics, engineering, chemistry, and computer science, to name a few. To prepare students for such opportunities, it is important to give them a strong foundation in the basics of quantum information science, in which quantum computing plays a central role. In this study, we discuss the development, validation, and evaluation of a tutorial on the Bloch sphere, a useful visual tool for developing intuition about single quantum bits (qubits), which are the basic building block of any quantum computer. Students' understanding was evaluated after they received traditional lecture-based instruction on the requisite topics, and again after engaging with the tutorial. We observe, analyze, and discuss their improvement in performance on concepts covered in the tutorial.

**Introduction**

Quantum information science and engineering (QISE) is an exciting interdisciplinary field that has applications in quantum computing, quantum communication and networking, and quantum sensing, which are attractive to scientists and engineers for many reasons. Computer scientists and engineers are developing quantum algorithms for various problems, including ones that become impractical for classical computers to solve at large scales. For example, on a classical computer, the problem of factoring products of large prime numbers scales exponentially with the size of the prime numbers, but on a quantum computer utilizing Shor's algorithm, the problem scales roughly as a polynomial instead. For future applications in science, physicists and chemists are also excited about the potential of quantum computers to solve important problems in their disciplines in which solving the Schrödinger equation plays an important role. The development of robust quantum bits (qubits) and scalable quantum computers demands the expertise of physicists and engineers alike. For all these reasons and more, this area of study holds a great amount of promise for students from many science and engineering disciplines who are interested in careers in QISE-related fields [1,2].

One teaching tool for the introduction of quantum states and their visualization is the Bloch sphere, which allows for visualization of the states of a qubit, the fundamental functional unit of a quantum computer. It can be an important and powerful aid for understanding the properties of two-state systems, but students often have difficulties with it. Additionally, the Bloch sphere is a highly useful tool for current research, including in quantum sensing and tomography, and experimentalists in the field routinely use it to characterize a single qubit in their work. The Bloch sphere allows one to graphically understand a single-qubit state, including mixed states via the density matrix, and the operations that can be done through single-qubit gates.

Here we describe the development, validation and implementation of a research-based tutorial on the Bloch sphere. The course in which the Bloch sphere tutorial was implemented is an interdisciplinary course involving students from many fields. The Bloch sphere is taught as a foundational topic in this course, which was the only mandatory course in a new "Foundations of Quantum Computing and Quantum Information" undergraduate certificate program at a large research university in the United States.

The Bloch sphere is a mathematical mapping of a complex two-dimensional Hilbert space onto the surface of a unit sphere in 3-D real space, which does not have any intrinsic mapping onto 3-D physical space. The Bloch sphere brings an additional geometric interpretation to what is otherwise a linear algebra-heavy topic, which is valuable in providing more context, additional perspectives, and intuitive representations to the concept of a qubit. States that require three independent real numbers to describe in two-dimensional Hilbert space, if one considers normalization (i.e., $(a_1 + ia_2)|0\rangle + (a_3 + ia_4)|1\rangle$, where $\sum a_i^2 = 1$, where $a_i$ are real numbers), only need two to be located on the Bloch sphere: the angles $\theta$ and $\phi$. This is accomplished by re-phasing any state such that the $|0\rangle$ component is real and non-negative; such a state can be expressed in the form $\cos\frac{\theta}{2}|0\rangle + \sin\frac{\theta}{2}e^{i\phi}|1\rangle$. This Bloch sphere tutorial helps students learn the definitions, conventions, and parameters that are used to construct the Bloch sphere, and after laying down the basics, it helps students explore various visual interpretations of the Bloch sphere. Points on the surface of the Bloch sphere represent pure states, while points inside the Bloch sphere represent mixed states; though mixed states are outside the scope of our targeted introductory level, familiarity with this basic Bloch sphere representation that we focus on in the tutorial provides a foundation upon which additional concepts can be built in future instruction.

Prior research suggests teaching quantum mechanics (QM) can often result in common difficulties shared by a number of students, and moreover, that these difficulties can be mitigated or eliminated by well-designed research-validated learning tools [3]. For instance, others have worked on developing interactive activities [4–6], investigating and leveraging student models [7–9], developing visualizations [10–12], and finding pedagogical value in authentic problems [13,14]. Other investigations have been made into student difficulties on a general level [15–19], on specific topics [9,20], in undergraduate and graduate contexts [21–24], and by considering an epistemological perspective [25]. Difficulties include the basic formalism [18,26], notation [27], wavefunctions [19,28], the concept of probability [29], measurement [19,26,30,31], and transferring learning from one context to other contexts [17,32]. We have been researching student difficulties and using the research to guide the development of learning tools for concepts covered in undergraduate QM courses [33]. Previous work from our group includes Quantum Interactive Learning Tutorials (QuILTs) on topics such as the Mach-Zehnder interferometer and quantum key distribution, and Clicker Question Sequences (CQSs) on topics such as the basics and change of basis, quantum measurement, time-development, and measurement uncertainty of two-state quantum systems [34–39]. Even after traditional lecture-based instruction, students may not yet have a strong grasp of important concepts, but further engagement and practice using a research-validated tutorial may help them develop additional fluency with those concepts. To that end, we have developed and validated a tutorial to help students learn about the Bloch sphere.

Given the interdisciplinary nature of the field of quantum information science, for which this tutorial is intended, there is a need to standardize the language to be accessible and unambiguous for everyone regardless of background. Some trends have already taken hold in the field, including a distinction between interpreting phenomena "classically" as opposed to "quantumly," rather than "quantum mechanically"; speaking of "measuring qubits" as opposed to measuring a physical observable; referring to a "measurement basis"; or measuring specific states such as $|0\rangle$ or $|1\rangle$ as outcomes rather than the corresponding eigenvalues. These are linguistic constructions that quantum physicists have typically not used. However, as there is a one-to-one correspondence, e.g., between the eigenvalues obtained by making a measurement of an observable in a specific state, and the eigenstates that are represented in the standard basis as $|0\rangle$ and $|1\rangle$, it can serve as useful shorthand to say that a qubit is measured to be in the $|1\rangle$ state to convey that a measurement made on the system yields the eigenvalue corresponding to the $|1\rangle$ state.

In this research on the development, validation, and implementation of the Bloch sphere tutorial, we opt to use the prevailing terminology in the field so that students can become familiar with the language used by their textbooks, instructors, and other professionals in their studies. Not coincidentally, this language also serves to express many concepts more directly and in fewer words, and thus may reduce the cognitive load needed to learn them, as many of the physical details are immaterial to the contexts in which these students will apply the concepts that they learn. All that said, it is still crucial to maintain the integrity of language used to educate physicists who will be working to develop robust qubits and build real quantum computers, work that very much requires understanding the entire quantum physics taught in typical undergraduate and graduate physics courses.

**Methodology**

*Development and validation*

Student difficulties and potential responses to questions were explored through preliminary open-ended questions asked on an exam in an undergraduate QM course following traditional lecture-based instruction on the requisite concepts. Common student difficulties were identified based on these student responses. These difficulties were then used as a guide over the subsequent development of the tutorial. Four students were interviewed several times using a think-aloud protocol, spanning a total of roughly fifteen hours. Their feedback was used to gauge overall flow and whether the tutorial was at the appropriate level, as well as to identify blind spots, and their suggestions were incorporated into the subsequent versions of the tutorial. Throughout, the tutorial was repeatedly iterated with constant discussions among the authors, and with three additional faculty experienced in teaching QM and solid-state physics continually contributing feedback as well.

Of particular note are the illustrations developed for the tutorial, as the Bloch sphere is predominantly a visual tool. It was of critical importance to represent the Bloch sphere and Cartesian axes without significant distortion. For this, a method resembling isometric projection was selected, which offsets all three axes from a straight-on view by a roughly equal amount. This method introduces only minor compromises in perspective that were considered reasonable.

Additional care was taken to color-code the Cartesian axes while ensuring that they remained distinguishable even in black and white without becoming too light to see. Efforts were also made to have the diagram read clearly as a sphere at an initial glance, using suggestive shading and guideline cues. These visualizations are supplemented by fully accurate, computer-generated graphics from a tool independently developed by the University of St. Andrews as part of the QuVis project [4], which is supported by IOP, which students were directed to use to verify their predictions and answers to some questions in the tutorial. With these two types of diagrams working in tandem, it is hoped that students would be able to mentally represent the structure and develop cognitive fluency with its geometry. The tutorial can be found in the accompanying supplementary data (https://iopscience.iop.org/article/10.1088/1361-6404/ad2393/data).

Developed and validated alongside the tutorial were a pre-test and post-test to evaluate students' understanding of the underlying concepts. The two tests contained isomorphic questions with minor changes to some details such as angles or given states. The post-test versions of these questions are provided in the Appendix.

*Learning objectives*

The tutorial consists of three sections, with broad learning objectives as follows:
1. *Construction:* Students should be able to describe how any single-qubit state can be transformed into the form $\cos\frac{\theta}{2}|0\rangle + \sin\frac{\theta}{2}e^{i\phi}|1\rangle$, and depict such a state on the Bloch sphere.
2. *Measurements:* Students should be able to identify the outcomes of measurements in a given state when a measurement basis is provided, qualitatively describe the probability of obtaining each outcome using visual cues, and explicitly calculate the probability of obtaining a particular outcome when given enough information regarding angles.
3. *Geometric intuition:* Students should be able to qualitatively describe and diagrammatically indicate answers to various questions, such as the set of all possible states that can yield a particular probability distribution when measured in a given basis.

*Course implementation*

The tutorial was administered at a large research university in the United States, in a multi-disciplinary undergraduate course titled "Foundations of Quantum Computing and Quantum Information." As the name suggests, the course focuses on an introduction to quantum information and (particularly) quantum computing. Aimed at undergraduate students from all science and engineering majors, it is the only mandatory course for a Quantum Computing and Quantum Information certificate available to interested undergraduate students across disciplines. The class comprised 28 students of diverse backgrounds: seven from engineering science, computer engineering, or industrial engineering; seven from computer science; six from math; two from chemistry; and twelve from physics. Some students chose more than one major from these disciplines. Most of them were sophomores, juniors, and seniors (one first-year student from the College of General Studies was enrolled). The more coarse-grained structure of the course and lack of pre-requisite knowledge related to quantum information enables students of any background to benefit from taking it. The only prerequisites are Calculus I and II, which serve as a proxy to determine students' ability to engage with the math, predominantly linear algebra, taught in the course in a self-contained way. The curriculum focused on ideas such as two-state systems, quantum gates, and doing measurements, but was less concerned with the details of making a good qubit, correcting errors, and the technical physics behind gating and measurements. Selected as the course text was Thomas Wong's *Introduction to Classical and Quantum Computing*.

The Bloch sphere tutorial was implemented after traditional lecture-based instruction on the relevant concepts. The students were first given a pre-test to establish a baseline level of knowledge following the lectures and traditional homework, and then they were assigned the tutorial for homework. Afterwards, once students submitted the homework, they were given the post-test. Common difficulties and performance improvements are discussed in the following sections. Two researchers graded a fifth of the pre- and post-tests. After discussion, they converged on a rubric for which the inter-rater reliability was greater than 90%. Following this, one researcher graded the remaining pre- and post-tests. Each student response to an open-ended question was graded on a three-tiered scale of 0, 0.5, or 1 point, for each salient unit of a problem. (A few problems had multiple aspects that were graded in this way, such as one that asked students to identify two angles.)

**Results: Student difficulties**

*Overall phase vs. relative phase*

Many students were not especially clear on the identification of quantum states that are equivalent after traditional lecture-based instruction. Equivalent quantum states are described in the tutorial as states that yield the same outcomes in all measurement bases, and in individual interviews, students had no difficulty interpreting the intended meaning of these terms. To evaluate students on their knowledge, question 1 on the pre- and post-test (see Appendix) posed a scenario in which there was a pair of states that differed in overall phase, and one with a pair that differed in relative phase, asking if each pair consisted of equivalent states. On the pre-test, most students answered yes or no for both pairs of states, making no distinction between the two pairs, making the overall correctness rate for the question close to 50% (see Table 1). A few stated "no" for the equivalent pair and "yes" for the nonequivalent pair, without indicating particularly compelling answers for why they provided these responses. Overall, it appears that many students were confused.

The tutorial helps students learn that two states that differ by an overall phase yield outcomes with identical probabilities (in any basis), while two states with different relative phases, which are not equivalent quantum states, may yield outcomes with the same probabilities in one basis but not another. To do this, students were given various pairs of states in the standard basis $\{|0\rangle, |1\rangle\}$, and were instructed to calculate the probabilities of measuring each outcome in both the $\{|0\rangle, |1\rangle\}$ basis and the $\{|+\rangle, |-\rangle\}$ basis, where $|+\rangle = \frac{1}{\sqrt{2}}\big(|0\rangle + |1\rangle\big)$ and $|-\rangle = \frac{1}{\sqrt{2}}\big(|0\rangle - |1\rangle\big)$.

Students performed better on the post-test for this concept, but there is still room for improvement, especially when compared to the other concepts (see Table 1). One possible reason is that the cues may have been more subtle for this question than others (e.g., there were no obvious diagrams that could evoke particular concepts from the tutorial). It is also possible that some students were getting too focused on the details of the calculations that were asked of them in the tutorial, and that the cognitive load distracted them from the important conclusion that equivalent states only differ by an overall phase. The illustration used in the tutorial could thus be improved; for example, by providing some of the basis change steps in advance to reduce the effort required to understand the problem statement, or having different hypothetical students make correct and incorrect statements in a written discussion, and asking students to reflect upon the validity of each.

*Definitions of $\theta$ and $\phi$ on the Bloch sphere*

On pretest questions 3a-d, most students correctly associated the angle in the argument of the cosine and sine functions in a state $\cos\frac{\theta}{2}|0\rangle + \sin\frac{\theta}{2}e^{i\phi}|1\rangle$ with the polar angle. However, they frequently neglected to multiply $\frac{\theta}{2}$ by 2 to obtain $\theta$, resulting in them drawing a polar angle half the required size on the Bloch sphere. The other type of common incorrect answer centered around not starting the angles $\theta$ and $\phi$ from their conventional starting points, the *z*-axis and *x*-axis, respectively. As a further example of this, on a cross-section of the Bloch sphere in questions 5a-b, some students on the pre-test considered the state overlapping with the $-z$-axis to be the $|0\rangle$ state, and the one overlapping with the $+z$-axis to be the $|1\rangle$ state, e.g., because of an association of the "lower" state with a lower energy. This is the reverse of the typical convention, and some of these students' responses were inconsistent between questions, since they had used the typical convention to answer prior questions.

The tutorial calls explicit attention to the $\frac{\theta}{2}$ present in the form $\cos\frac{\theta}{2}|0\rangle + \sin\frac{\theta}{2}e^{i\phi}|1\rangle$, and included an exercise directing students to find the values of the angles for states that corresponded to the Cartesian axes. Both issues discussed above were largely corrected on the post-test, while students were given at least half credit if they explicitly stated the correct values for the angles. It was interesting, however, that on post-test question 3a, some students who explicitly wrote down the correct value of $\phi$ (which in this case was $\frac{2\pi}{3}$), when drawing the angle on the Bloch sphere, did not extend it past the *y*-axis as they should have, affecting their answer to question 3c. This could be due to a careless mistake, but these students who then used their diagram to answer question 3c did so correctly with respect to their diagram.

One last difficulty observed among multiple students, primarily on the post-test because the other difficulties had greatly diminished, was the labeling of not only $\theta$ (correctly) as the polar angle with respect to the $+z$-axis, but also of $\phi$ (incorrectly) as a polar angle measured with respect to the $+x$-axis rather than an azimuthal angle in the *x-y* plane. The difference between the two (depicted in Figure 1) is rather subtle, and the tutorial does not go into particularly lengthy detail about the definitions of "polar" and "azimuthal" beyond illustrating the terms with some diagrams.

**Figure 1.** [Left] A conventional depiction in physics of the angle $\phi$, defined as the azimuthal angle in the *x-y* plane. [Right] A non-conventional depiction provided by some students,

appearing to interpret $\phi$ as a polar angle with respect to the positive $x$-axis (similarly to $\theta$, not shown, which is defined with respect to the positive $z$-axis).

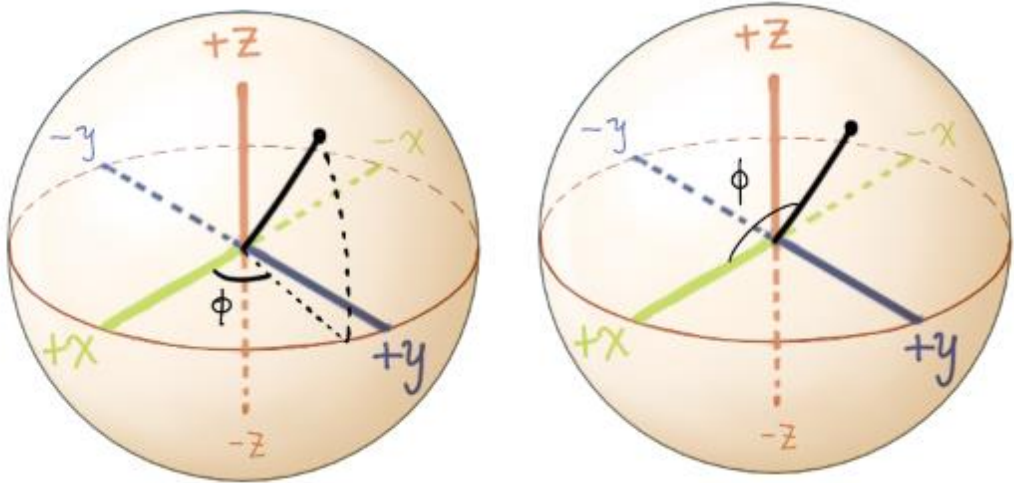

*Difficulties with the Born rule*

Question 3b on the pre-test and post-test asked students to provide the possible measurement outcomes for the qubit in the standard basis $\{|0\rangle, |1\rangle\}$, and their respective probabilities. On the pre-test, students tended to provide dichotomous answers that were either completely correct or not on the right track. The most common difficulty that appeared with some consistency was directly squaring the complex exponential rather than squaring its modulus, though only a few students did this. Only this particular difficulty was present on the post-test as well, and the others had been completely addressed.

*Measurements made in bases other than the standard basis*

The tutorial has a section dealing with measurements made in arbitrary bases, when the polar angle that the state makes with respect to the measurement basis states is known. In this case, if $\theta$ is the angle between the given state and one of the basis states, the probability that the measurement yields that basis state is $\cos^2 \frac{\theta}{2}$. This knowledge was evaluated by pre- and post-test questions 4a-c.

On the pre-test, many students gave multiple incorrect answers. After engaging with the tutorial, the vast majority of students gave answers in the correct form, with some only forgetting to divide the appropriate $\theta$ by 2, for which they were given partial credit.

*Identifying states for which a particular measurement outcome is more likely*

On questions 5a-c on the pre-test and post-test, students were given a cross-section of the Bloch sphere (i.e., a circle) containing the $x$-axis and the $z$-axis, corresponding to the $\{|+\rangle, |-\rangle\}$ and $\{|0\rangle, |1\rangle\}$ bases, respectively. They were asked to indicate states that had a greater chance of yielding one of the outcomes than the other when measured in each basis. Expected answers are states on the edge of the circle positioned close to, or even directly on top of, the states in question, e.g., a state with a greater chance of yielding $|0\rangle$ than $|1\rangle$ when measured in the $\{|0\rangle, |1\rangle\}$ basis is any state on the upper half of the cross-section shown as a circle.

On the pre-test and the post-test, the most common difficulty for questions 5a-c was for students to indicate states on the interior of the cross-section of the Bloch sphere, rather than the edge of the circle. Such states are interpreted as mixed states, which were beyond the scope of the tutorial and the course, but to the extent that students' indicated states were closer to the correct basis state, such responses were given partial credit. This difficulty was observed in lower numbers on the post-test, but was not eliminated entirely. One student initially answered with points on the edge of the circle, before erasing these answers and instead choosing points along the axes on the interior of the cross-section. It is possible that at least some students intended such points to represent locations on the surface of the Bloch sphere some distance above or below the page, in which case they would be acceptable answers. However, this remains unclear (these types of answers and reasoning did not come up in interviews before the in-class implementation), and it is also possible that some students had a strong preference for remaining along the axes that represent a measurement basis. They may think, for example, that only states located along the $\pm z$-axes may be measured with what they consider "valid" outcomes in the $\{|0\rangle, |1\rangle\}$ basis.

On the pre-test, it was also relatively common for students to mislocate the $|0\rangle, |1\rangle, |+\rangle$, and $|-\rangle$ states. For some students, this was as simple as switching the $|0\rangle$ and $|1\rangle$ states, but others appeared to think that the $|0\rangle$ and $|1\rangle$ states were located on the positive and negative $x$-axis, in effect rotating the Bloch sphere by a quarter turn counterclockwise. These difficulties appeared to be related only to an initial unfamiliarity with the conventions and definitions of the Bloch sphere, and had disappeared on the post-test.

In the preliminary investigation of difficulties, it was somewhat common for students to explain that the $|0\rangle$ state lies on the positive $z$-axis while the $|1\rangle$ state lies on the positive $x$-axis, which is incorrect. This difficulty likely comes from the states $|0\rangle$ and $|1\rangle$ being referred to as "orthogonal" to one another, and such orthogonality of these states does manifest in a right angle between them in two-dimensional Hilbert space. However, in the mapping onto the Bloch sphere, orthogonal states appear diametrically opposite, not at right angles, to each other, and the tutorial emphasizes this point. While in this implementation, only one student on the pre-test provided an answer to this effect, this difficulty is likely to come up when students receive instruction on the Bloch sphere.

*Measurements for which the outcome is certain*

On questions 6 and 7 on the pre-test, many students did not convey that a qubit would yield an outcome with 100% certainty if and only if the qubit was in one of the measurement basis states. This was true whether they were asked to identify states for which a measurement outcome is certain (question 6), or the number of bases in which a given state could be measured to yield a certain outcome (question 7, in which four possible answers were given: zero, one, two, or infinitely many). On the pre-test, many students left question 6 blank, and all answers in the multiple-choice question 7 were observed; one or two bases were most popular, but at least two students each chose zero or infinitely many bases. (The "zero bases" option was included with the idea that students may mistakenly think that since the state $|p\rangle$ cannot be measured in the standard basis with 100% certainty, then neither can it be in *any* basis.) This appears to be challenging after traditional lecture-based instruction alone; however, on the post-test after the tutorial, nearly all students answered question 6 correctly, and much improvement was observed on question 7 as well, with three fourths providing the correct response (see Table 1).

For question 7, the answer spread was narrowed down to one basis (correct) or two bases. While most students who gave the latter answer did so with no elaboration, a few went on to explicitly give the two bases as ones consisting of the same two states, but written in reverse (e.g., $\{|p\rangle, |-p\rangle\}$ and $\{|-p\rangle, |p\rangle\}$), while noting that the two are functionally the same basis. This explanation was given full credit, and it is possible that some of the students who selected two bases as the answer to question 7 had this in mind. However, in interviews, some students were observed referring to bases by a single state, e.g., "the $|+\rangle$ basis" or "the $|p\rangle$ basis," so it may also be the case that some students did not have yet a fully developed idea of what constitutes a basis.

*Finding the set of all states given a probability distribution*

One other visual affordance offered by the Bloch sphere is the ability to narrow down the possible states a qubit is in after the probability distribution in one or more measurement bases is known. In the case of a qubit that yields $|0\rangle$ with 60% probability and $|1\rangle$ with 40% probability, all possible states that have this property are found on a circle on the surface of the Bloch sphere with $\cos^2\frac{\theta}{2} = 0.6$. Visually, this looks like a circle of constant "latitude" located slightly above the "equator" (i.e., the circle that intersects the *xy*-plane) of the Bloch sphere. Similar conclusions can be reached given the results of measurements made in any basis.

This was not something that most students were very familiar with immediately following traditional lecture-based instruction, as their answers on question 8 on the pre-test had no consistent underlying patterns. However, it appears that the tutorial was helpful in helping them learn this concept, as evidenced by the exceptional normalized gain and effect size yielding a final performance of 86% on the post-test (see Table 1). The tutorial dedicates a section to this visual representation, including questions that made use of the QuVis Bloch sphere visualization [4]. Students were asked to provide qualitative descriptions of what happens to a state on the Bloch sphere while varying either $\theta$ or $\phi$ while holding the other constant, and their interaction with this tool along with subsequent questions prodding them draw further conclusions appear to have been effective at helping students learn this concept.

Since some students did not appear to be clear about the distinction between an overall and relative phase, there could be a further opportunity to reinforce the visual meaning of a relative phase with this concept. For example, one could have them contemplate that knowing the measurement probabilities of each outcome fixes an angle $\theta$ with respect to the measurement basis, but varying the value of $\phi$ with respect to the basis (while holding $\theta$ constant) sweeps out a circle on the surface of the Bloch sphere.

A summary of the difficulties discussed in this section can be found in Table 2.

**Table 1.** Comparison of scores before and after the administration of the tutorial, along with normalized gain [40] and effect size as measured by Cohen's $d$ [41], for students who engaged with the CQS ($N = 28$).

|    | Pre-test mean | Post-test mean | Normalized gain | Effect size |
|----|---------------|----------------|-----------------|-------------|
| 1  | 66%           | 73%            | 0.21            | 0.11        |
| 2  | 100%          | 96%            | -               | -           |
| 3a | 57%           | 85%            | 0.65            | 0.91        |
| 3b | 66%           | 95%            | 0.84            | 0.87        |
| 3c | 93%           | 96%            | 0.50            | 0.16        |
| 3d | 82%           | 100%           | 1.00            | 0.66        |
| 4a | 38%           | 89%            | 0.83            | 0.74        |
| 4b | 36%           | 82%            | 0.72            | 0.64        |
| 5a | 50%           | 86%            | 0.71            | 0.98        |
| 5b | 50%           | 84%            | 0.68            | 0.90        |
| 5c | 66%           | 84%            | 0.53            | 0.52        |
| 6  | 55%           | 93%            | 0.84            | 0.99        |
| 7  | 39%           | 75%            | 0.59            | 0.77        |
| 8  | 29%           | 86%            | 0.80            | 1.55        |

**Table 2.** Summary of the difficulties described and student improvement on each between the pre-test and post-test.

| Difficulties | Pre-/post-test # \| comments |
|---|---|
| Difficulties with identifying equivalent states using overall or relative phase | 1 \| Some improvement |
| Every possible state can be plotted on the Bloch sphere with unique $\theta$ and $\phi$ | 2 \| High pre-test performance |
| Unfamiliarity with Bloch sphere conventions, e.g., which states are defined as $|0\rangle$ or $|1\rangle$; or the angles $\theta$ and $\phi$ for a given quantum state | 3a, 5a \| Major improvement |
| Difficulty identifying possible outcomes and calculating probabilities of those outcomes when measuring a qubit | 3b \| Major improvement |
| Determining which outcome is more likely given a state and a measurement basis | 3c, 3d \| Some improvement |
| Inability to interpret probability of measuring a particular outcome as a function of the polar angle between the given state and measurement basis state | 4a, 4b \| Major improvement |
| Difficulties identifying possible states that are more likely to yield one outcome than another | 5a, 5b, 5c \| Major improvement, though a number of students tended |
|  | to plot points on the interior of the Bloch sphere |
| Not knowing that a quantum state which is a measurement basis state can be measured with 100% certainty | 6, 7 \| Major improvement |
| Not knowing the set of all states that can be measured with the same probability distribution in a particular basis | 8 \| Major improvement |

**Discussion**

Students who engaged with the tutorial generally performed very well on the post-test. Only two students did not submit a complete tutorial by the deadline, and compared to their peers who did, they exhibited qualitative differences in their responses in addition to performing substantially worse on the post-test. It is possible that these differences are due to the fact that those two students did not engage with the Bloch sphere tutorial.

Physics majors and non-physics majors did not display appreciable differences in performance on the pre- and post-test. This may not be especially surprising, as the Bloch sphere is not typically discussed in detail in QM courses in a typical physics program.

However, especially on the pre-test, the geometry of spherical coordinates proved difficult for many students, which may be because spherical coordinates are not broadly covered in different science and engineering disciplines. That said, the tutorial greatly increased students' facility with spherical coordinate notations and conventions, as seen in improvement in questions 3a-d (see Table 1). In this, physics majors did appear to be more comfortable than non-physics majors, which suggests that their previous physics courses had introduced them to spherical coordinates.

Overall, findings from the pre- and post-tests suggest that the only concept that may need some refinement is the one evaluated by question 1, which asks about equivalent quantum states. (Question 7 has a similar post-test score, but much higher normalized gain and effect size from the lower pre-test score—see Table 1.) We are considering approaches to help students engage with this concept more succinctly and reduce their cognitive overload, as well as to encourage students to make connections back to this concept at several different relevant points in the tutorial.

**Ethical statement**

**Acknowledgments**

We thank the NSF for awards PHY-1806691 and PHY-2309260. RM is supported by the NSF under award DMR-1848336. We thank all students whose data were analyzed and Dr. Robert P. Devaty for his constructive feedback on the manuscript.

**Appendix**

Notes:
- In the Bloch sphere diagrams of this section, the $\hat{\mathbf{x}}, \hat{\mathbf{y}}$ and $\hat{\mathbf{z}}$ unit vectors point toward the viewer, while the $-\hat{\mathbf{x}}, -\hat{\mathbf{y}}$ and $-\hat{\mathbf{z}}$ unit vectors point away from the viewer.
- Solid lines indicate states on the front hemisphere (which point toward the viewer), while dotted lines indicate states on the back hemisphere.
- When a measurement basis is indicated, the larger ket label indicates the closer state (which points toward the viewer), while the smaller ket label indicates the state further away.

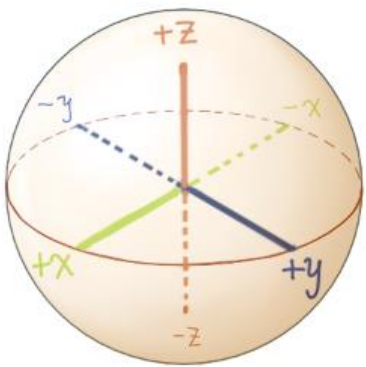

- In these questions, states are said to be "equivalent" if they yield identical measurement outcomes with identical probabilities for all observables (i.e., in all measurement bases).
- A state $|q\rangle = \cos\frac{\theta}{2}|0\rangle + \sin\frac{\theta}{2}e^{i\phi}|1\rangle$ can be found on the Bloch sphere at the position of the Cartesian vector $\mathbf{r} = \sin\theta\cos\phi\,\hat{\mathbf{x}} + \sin\theta\sin\phi\,\hat{\mathbf{y}} + \cos\theta\,\hat{\mathbf{z}}$, where $0 \leq \theta \leq \pi$ and $0 \leq \phi < 2\pi$.
  - The polar angle $\theta$ begins from the positive $z$-axis and sweeps toward the equator, and the azimuthal angle $\phi$ begins from the positive $x$-axis and sweeps counterclockwise about the $z$-axis.

**Warm-up**
Learning objectives: Students should be able to identify that multiplying by an overall phase does not change the state, but modifying a relative phase between the basis states does

1. Consider whether the following pairs of states are equivalent, i.e., states that yield identical measurement outcomes with identical probabilities in all bases:
   (A) $|q\rangle = \frac{\sqrt{3}}{2}|0\rangle + \frac{1}{2}|1\rangle$ and $|p\rangle = \frac{i\sqrt{3}}{2}|0\rangle + \frac{i}{2}|1\rangle$. Are $|q\rangle$ and $|p\rangle$ equivalent states?
   (B) $|b\rangle = \frac{1}{\sqrt{2}}|0\rangle + \frac{i}{\sqrt{2}}|1\rangle$ and $|d\rangle = \frac{1}{\sqrt{2}}|0\rangle - \frac{i}{\sqrt{2}}|1\rangle$. Are $|b\rangle$ and $|d\rangle$ equivalent states?
2. True or false: Any normalized state written as $a|0\rangle + b|1\rangle$ corresponds to a unique point on the Bloch sphere.

**Measurements**
Learning objectives: Students should be able to indicate a state on the Bloch sphere given $\theta$ and $\phi$; identify the outcomes of measurement when presented with a generic state $|q\rangle$ on the Bloch sphere with $\{|0\rangle, |1\rangle\}$ being the measurement basis; describe the measurement outcomes and calculate the probabilities of those outcomes for the given generic state $|q\rangle$ when $\{|0\rangle, |1\rangle\}, \{|+\rangle, |-\rangle\}, \{|+i\rangle, |-i\rangle\}$ are each used as measurement bases (by changing basis)

3. Consider the state $|q\rangle = \cos\frac{\pi}{8}|0\rangle + \sin\frac{\pi}{8} e^{i\frac{2\pi}{3}}|1\rangle$.

   (A) On the Bloch sphere below, indicate the state $|q\rangle$. Label $\theta$ and $\phi$.

   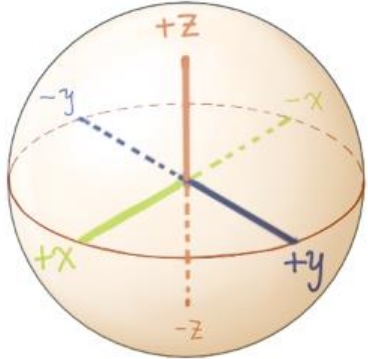

   (B) The state $|q\rangle = \cos\frac{\pi}{8}|0\rangle + \sin\frac{\pi}{8} e^{i\frac{2\pi}{3}}|1\rangle$ is measured in the $\{|0\rangle, |1\rangle\}$ basis. What are the outcomes? What is the probability of yielding each outcome? (There is no need to find numerical values.)

   (C) The state $|q\rangle$ is measured in the $\{|+\rangle, |-\rangle\}$ basis. Which outcome is more probable?

   (D) The state $|q\rangle$ is measured in the $\{|+i\rangle, |-i\rangle\}$ basis. Which outcome is more probable?

4. Consider a qubit in the state $|q\rangle = a|0\rangle + b|1\rangle$ shown below.

   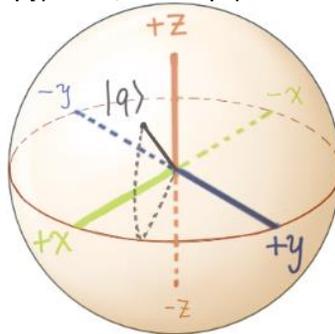

   $|q\rangle$ is illustrated in the following three diagrams, each one highlighting a different measurement basis. The (nonstandard) notations $\theta_1, \theta_2$, and $\theta_3$ will be used to refer to the angles made with respect to one of the states in this measurement bases.

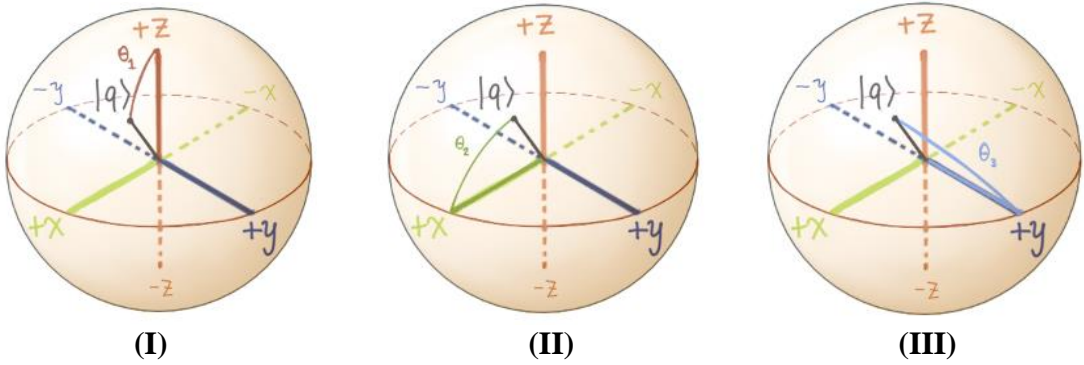

(A) In diagram (I), $|q\rangle$ makes an angle $\theta_1$ with the $|0\rangle$ state.
(B) In diagram (II), $|q\rangle$ makes an angle $\theta_2$ with the $|+\rangle$ state.
(C) In diagram (III), $|q\rangle$ makes an angle $\theta_3$ with the $|+i\rangle$ state.

Express your answers to (A) and (B) in terms of $\theta_1$, $\theta_2$, and $\theta_3$.

(A) When $|q\rangle$ is measured in the $\{|+\rangle, |-\rangle\}$ basis, what is the probability that $|q\rangle$ will collapse into the state $|+\rangle$?

(B) When $|q\rangle$ is measured in the $\{|+i\rangle, |-i\rangle\}$ basis, what is the probability that $|q\rangle$ will collapse into the state $|-i\rangle$? (Note: **NOT** $|+i\rangle$!)

5. Below is shown a cross-section of the Bloch sphere, with the axes indicated.

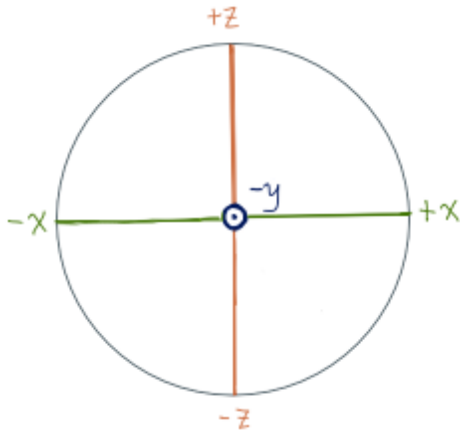

(A) On the cross-section of the Bloch sphere, indicate a state of your choosing that, when measured in the $\{|0\rangle, |1\rangle\}$ basis, has a higher chance of yielding $|0\rangle$ than $|1\rangle$.

(B) On the cross-section of the Bloch sphere, indicate a state of your choosing that, when measured in the $\{|0\rangle, |1\rangle\}$ basis, has a higher chance of yielding $|1\rangle$ than $|0\rangle$.

(C) On the cross-section of the Bloch sphere, indicate a state of your choosing that, when measured in the $\{|+\rangle, |-\rangle\}$ basis, has a higher chance of yielding $|+\rangle$ than $|-\rangle$.

**Geometric intuition**
Learning objectives: Students should be able to identify that every basis has only two orthonormal states that can yield results with 100% certainty; describe that no two distinct measurement bases on the Bloch sphere are compatible (an eigenstate of one basis cannot be an eigenstate of any other basis); identify that a circle represents the set of all states that a single measurement basis can distinguish from all other states, since the measurement basis cannot determine relative phase in that basis

6. On the Bloch sphere below, the $\{|q\rangle, |-q\rangle\}$ basis is illustrated. Indicate (or otherwise specify) all the states in which a measurement in this basis yields a result with 100% certainty.

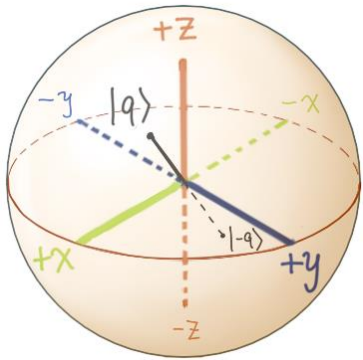

7. Consider the state $|p\rangle$ on the Bloch sphere below.

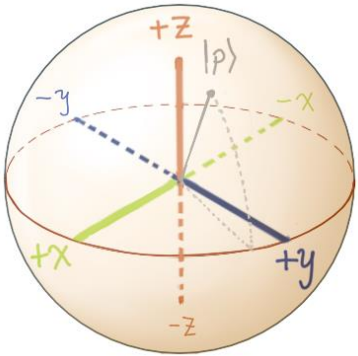

You are in state $|p\rangle$ and want to make a measurement such that you get a certain outcome with 100% probability. How many measurement bases can you choose to accomplish this?
  a. 0
  b. 1
  c. 2
  d. Infinitely many

8. On the Bloch sphere below, indicate the set of all states that, when measured in the illustrated $\{|q\rangle, |-q\rangle\}$ basis, will yield $|q\rangle$ with (approximately) 90% probability and $|-q\rangle$ with 10% probability.

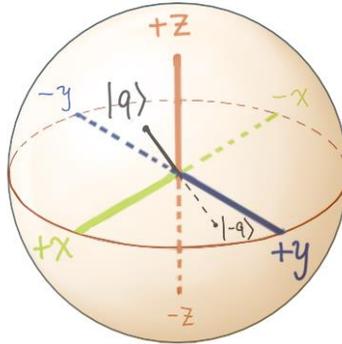